\documentclass[11pt]{article}

\usepackage[margin=1in]{geometry}
\usepackage{graphicx}
\usepackage{amsmath,amssymb,bm}
\usepackage{physics}
\usepackage{siunitx}
\usepackage{cite}
\usepackage{hyperref}
\hypersetup{colorlinks=true,linkcolor=blue,citecolor=blue,urlcolor=blue}

\newcommand{\splus}{\bm{s}^{+}}
\newcommand{\sminus}{\bm{s}^{-}}
\newcommand{\Smat}{\hat{\bm{S}}}
\newcommand{\Pin}{P_{\mathrm{in}}}
\newcommand{\Pout}{P_{\mathrm{out}}}
\newcommand{\Pnet}{P_{\mathrm{net}}}
\newcommand{\kz}{k_0}
\newcommand{\ii}{\mathrm{i}}

\title{\bf Coherently Assisted Wireless Power Transfer Through Poorly Transparent Barriers}
\author{Alex Krasnok\\
Department of Electrical and Computer Engineering,\\ Florida International University, Miami, Florida 33174, USA\\
\texttt{akrasnok@fiu.edu}}
\date{}

\begin{document}
\maketitle

\begin{abstract}
Poorly transparent barriers (e.g., reinforced walls, shielding panels, metallic or high-contrast dielectrics) strongly reflect incident radiation, limiting wireless power transfer (WPT) unless the barrier is structurally modified to support a narrowband transparency window. Here we introduce a barrier-agnostic alternative based on \emph{coherent scattering control}: a phase-locked auxiliary wave is launched from the receiver side with an amplitude and phase chosen from the measured complex scattering parameters of the barrier. In a two-port (single-channel-per-side) description, we derive closed-form conditions for (i) canceling back-reflection toward the transmitter and (ii) maximizing the \emph{net} extracted power at the receiver side. In the lossless limit these conditions imply unit transmitter-to-receiver efficiency (all transmitter power is routed to the receiver side) even when the barrier is nearly opaque under one-sided illumination. We validate the concept using (1) an analytically solvable high-index Fabry--P\'erot slab and (2) a numerically simulated perforated PEC metasurface exhibiting vanishing one-sided transmission; in both cases, coherent assistance yields near-unity transmission and large enhancement factors. We further analyze dissipative barriers using a receiver-side energy-balance metric, showing that substantial net delivery can persist well into the lossy regime. The approach is closely related to coherent perfect absorption and time-reversal ideas in wave physics, but targets \emph{reflectionless delivery through barriers} without modifying the obstacle itself.
\end{abstract}

\section{Introduction}

Wireless power transfer has progressed from early far-field visions \cite{Tesla1914} to modern near-field resonant schemes that enable efficient mid-range delivery \cite{Kurs2007,Karalis2008,Hui2014}. These advances underpin applications spanning consumer electronics, implants, and electric vehicles \cite{Sample2011,Kim2012,Ho2014PNAS,LiMi2015}. In many practical environments, however, energy must traverse barriers that are intentionally or unavoidably reflective (concrete walls with reinforcement, metallic enclosures, protective shields), so one-sided illumination yields poor transmission and strong standing-wave formation. Conventional remedies typically \emph{engineer the barrier} into a resonant ``window'' (e.g., subwavelength apertures or patterned metallic screens), leveraging extraordinary transmission effects \cite{Ebbesen1998,MartinMoreno2001,GarciaVidal2010,Hibbins2001,Suckling2005}. This structural approach is often narrowband and incompatible with existing infrastructure or security constraints.

This paper develops a different strategy: \emph{do not alter the barrier}; instead, shape the \emph{incident state} by injecting a phase-coherent auxiliary wave from the receiver side. Conceptually, this belongs to the broader program of coherent control of scattering and absorption \cite{Chong2010,Wan2011,Baranov2017NRM,Pichler2019Nature}, where multi-port excitation enables dramatic changes in a linear system response. Unlike coherent perfect absorption (CPA), which uses interference \emph{plus dissipation} to absorb all incident energy \cite{Chong2010,Wan2011}, here we use interference to \emph{cancel undesired scattering} (back-reflection) and \emph{route flux} to the receiver side. The method also connects to coherent-assisted impedance matching in WPT links \cite{Krasnok2018PRL,Krasnok2019Electronics} and is conceptually adjacent to active cancellation ideas employed in cloaking and wave manipulation \cite{Miller2006Cloak,AluEngheta2005,Sounas2015UnidirCloak,FleuryAlu2014Review}, though our focus is energy delivery through a barrier rather than invisibility.

\begin{figure}[t]
\centering
\includegraphics[width=\linewidth]{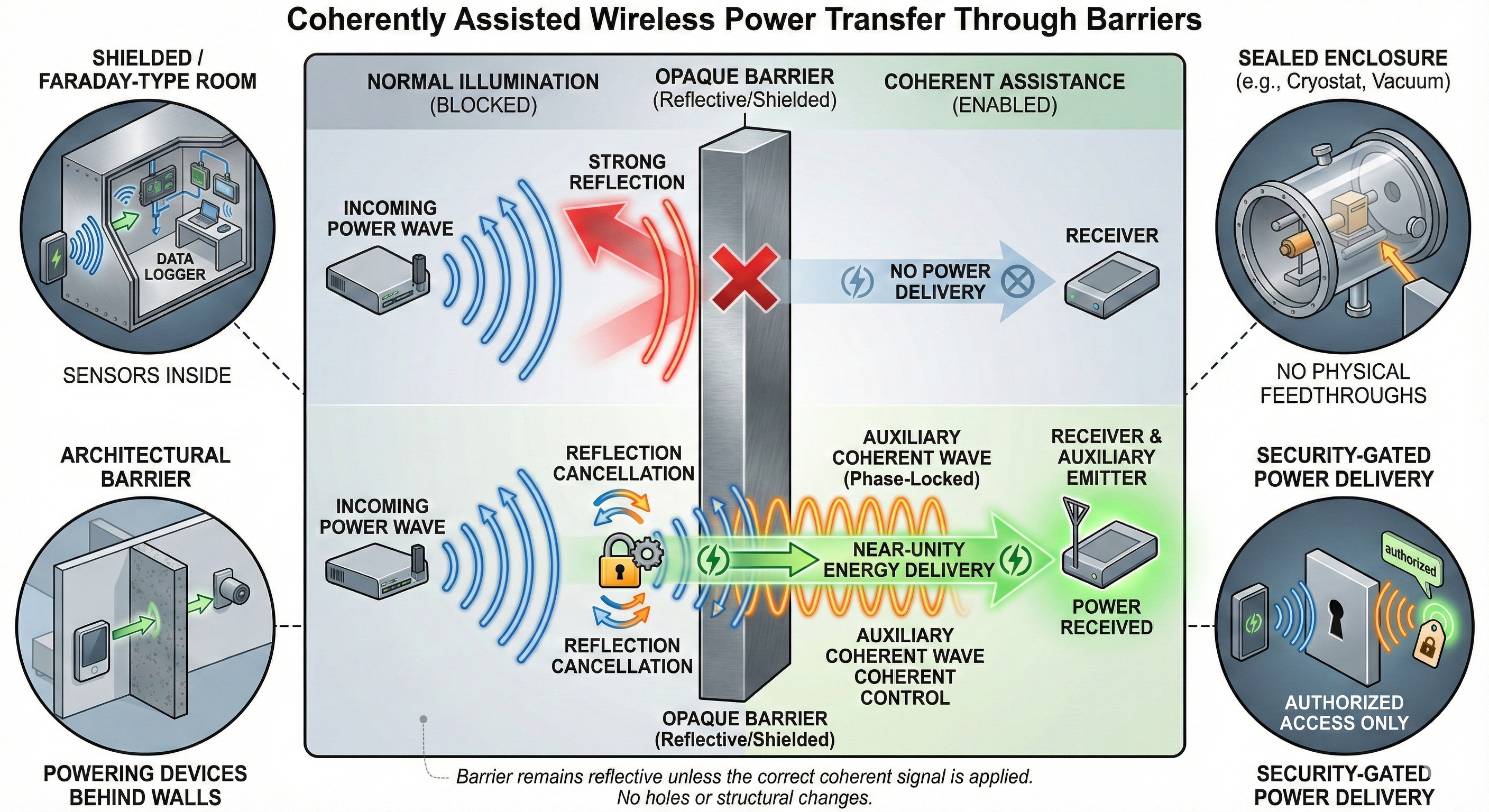}
\caption{\textbf{Concept.} Conventional WPT through a highly reflective barrier suffers strong back-reflection and poor transmission. In coherently assisted WPT, a phase-locked auxiliary wave is launched from the receiver side and tuned in amplitude and phase using the barrier's complex scattering parameters. In the lossless two-port limit, coherent assistance can cancel reflection toward the transmitter and route all transmitter power to the receiver side (unit transmitter-to-receiver efficiency). In lossy barriers, coherent assistance cannot eliminate dissipation but can still strongly increase the net received power.}
\label{fig:fig1}
\end{figure}

Figure~\ref{fig:fig1} summarizes the operating principle and motivates the target environments. Under conventional one-sided illumination, an opaque (reflective or shielded) barrier returns most of the incident power toward the source and blocks delivery, a situation that is well recognized in building and indoor/outdoor propagation models through large building-entry and penetration losses \cite{TR38901,ITURP2040,ITURP2346}. Coherent assistance activates a second excitation on the receiver side that is phase-locked to the transmitter and whose complex amplitude can be chosen from the measured scattering response of the barrier. The resulting two-sided scattering state suppresses the backward-propagating wave (reflection cancellation) and routes flux toward the receiver. In the lossless two-port limit this can yield unit transmitter-to-receiver efficiency; for dissipative and multi-channel barriers the same scattering framework predicts the maximum achievable net delivery and provides a direct recipe for waveform optimization.

The application panels in Fig.~\ref{fig:fig1} emphasize settings where modifying the barrier is undesirable or infeasible but receiver-side access exists. First, shielded rooms and Faraday-type enclosures are widely deployed to create electromagnetically isolated environments (e.g., EMC qualification, secure facilities, or low-noise measurements), with shielding performance characterized by standardized procedures such as IEEE~299 \cite{IEEE2992006}. Magnetic shielded rooms used for biomagnetism and precision measurements represent an extreme example of ``barriers by design,'' where external fields must be suppressed by orders of magnitude \cite{Bork2001MSR,Holmes2022MSR}. Second, sealed enclosures such as cryostats and UHV/vacuum systems routinely rely on specialized feedthroughs and wiring strategies, where vacuum integrity, cleanliness, thermal contraction, and wiring heat loads become dominant engineering constraints \cite{BalshawPracticalCryogenics,Cianciolo2018Feedthrough,Vilcins2014Feedthrough}. Third, architectural barriers and deep-indoor deployments explicitly confront penetration loss through reinforced concrete, metallized glass, and multi-layer walls \cite{TR38901,ITURP2040,ITURP2346}. These examples motivate a barrier-agnostic route to through-barrier power delivery based on coherent control rather than structural modification.

\section{Theory: coherent control of a two-port barrier}

\subsection{Scattering formulation and power normalization}

Consider a linear time-invariant barrier supporting two accessible scattering channels (``ports'') on the transmitter and receiver sides. Using power-wave normalization \cite{Pozar2012}, the incident and outgoing complex amplitudes are grouped into vectors
\begin{equation}
\splus =
\begin{bmatrix}
s_1^{+}\\ s_2^{+}
\end{bmatrix},
\qquad
\sminus =
\begin{bmatrix}
s_1^{-}\\ s_2^{-}
\end{bmatrix},
\qquad
\sminus = \Smat\, \splus,
\label{eq:s_scatter}
\end{equation}
where $|s_i^{\pm}|^2$ is proportional to power carried by the corresponding incoming/outgoing wave. For a reciprocal two-port barrier,
\begin{equation}
\Smat=
\begin{bmatrix}
r_{11} & t\\
t & r_{22}
\end{bmatrix},
\label{eq:S_twoport}
\end{equation}
with complex transmission $t$ and reflection coefficients $r_{11}$, $r_{22}$ defined with respect to the same reference impedances.

We define the total incident power and (selected) outgoing power as
\begin{equation}
\Pin = \|\splus\|^2 = |s_1^{+}|^2+|s_2^{+}|^2,
\qquad
\Pout^{(2)} = |s_2^{-}|^2.
\label{eq:PinPout}
\end{equation}
A convenient \emph{fractional routing} metric is then
\begin{equation}
\Sigma \equiv \frac{\Pout^{(2)}}{\Pin}
=\frac{|s_2^-|^2}{|s_1^{+}|^2+|s_2^{+}|^2}\ge 0.
\label{eq:Sigma_def}
\end{equation}
However, for WPT one must also track the fact that the auxiliary source at port~2 injects energy. The physically meaningful \emph{net extracted power} at the receiver side is
\begin{equation}
\Pnet^{(2)} \equiv |s_2^-|^2-|s_2^+|^2,
\label{eq:Pnet}
\end{equation}
and the transmitter-to-receiver efficiency is
\begin{equation}
\eta \equiv \frac{\Pnet^{(2)}}{|s_1^+|^2}.
\label{eq:eta_def}
\end{equation}
In the ideal case, $\eta=1$ means \emph{all transmitter-incident power is delivered to the receiver side} with zero net energy supplied by the auxiliary port beyond shaping the scattering state.

\subsection{Two-port coherent assistance with prescribed amplitude and phase}

Let the transmitter excitation be $s_1^{+}=1$ (unit amplitude at frequency $\omega$), and let the auxiliary wave be
\begin{equation}
s_2^{+} = a\,e^{\ii\varphi},\qquad a\ge 0,
\label{eq:aux_def}
\end{equation}
with the same temporal dependence $e^{\ii\omega t}$ suppressed. From Eq.~\eqref{eq:S_twoport},
\begin{equation}
s_2^- = t + r_{22}\,a e^{\ii\varphi},
\qquad
s_1^- = r_{11}+t\,a e^{\ii\varphi}.
\label{eq:outputs}
\end{equation}
Substituting into Eq.~\eqref{eq:Sigma_def} yields
\begin{equation}
\Sigma(a,\varphi)=
\frac{|t|^2+a^2|r_{22}|^2+2a\,\Re\!\left[t r_{22}^* e^{-\ii\varphi}\right]}{1+a^2}.
\label{eq:Sigma_a_phi}
\end{equation}
For fixed $a$, $\Sigma$ is maximized by choosing the phase to align the two contributions at the receiver-side output:
\begin{equation}
\varphi_{\Sigma,\mathrm{opt}}(a)=\arg\!\left(t r_{22}^*\right),
\qquad
\Sigma_{\max}(a)=\frac{\big(|t|+a|r_{22}|\big)^2}{1+a^2}.
\label{eq:Sigma_optphase}
\end{equation}
Maximizing $\Sigma_{\max}(a)$ over $a$ gives
\begin{equation}
a_{\Sigma,\mathrm{opt}}=\frac{|r_{22}|}{|t|},
\qquad
\Sigma_{\max}=|t|^2+|r_{22}|^2.
\label{eq:Sigma_optamp}
\end{equation}
In a \emph{lossless two-port} barrier, $\Smat$ is unitary and thus $|t|^2+|r_{22}|^2=1$, implying $\Sigma_{\max}=1$ even if $|t|^2\ll 1$ under one-sided illumination.

\subsection{Closed-form optimum for \emph{net} received power and reflection cancellation}

The more stringent WPT requirement is maximizing $\eta$ (net extracted power). Using Eqs.~\eqref{eq:Pnet} and \eqref{eq:outputs} with $s_1^+=1$ and $x\equiv s_2^+$,
\begin{equation}
\Pnet^{(2)}(x)=|t+r_{22}x|^2-|x|^2
=|t|^2 + 2\Re\!\left[t r_{22}^* x^*\right] + (|r_{22}|^2-1)|x|^2.
\label{eq:Pnet_quad}
\end{equation}
If the barrier is passive (no gain) and port~2 is power-normalized, typically $|r_{22}|<1$ so the quadratic is concave and has a unique maximizer:
\begin{equation}
x_{\eta,\mathrm{opt}}=\frac{t\,r_{22}^*}{1-|r_{22}|^2},
\label{eq:xopt_lossy}
\end{equation}
with $\eta_{\max}=\Pnet^{(2)}(x_{\eta,\mathrm{opt}})$.

In the \emph{lossless two-port} case, unitarity implies $1-|r_{22}|^2=|t|^2$, so Eq.~\eqref{eq:xopt_lossy} reduces to a remarkably simple condition,
\begin{equation}
s_{2,\mathrm{opt}}^+ = x_{\eta,\mathrm{opt}}=\frac{r_{22}^*}{t^*}.
\label{eq:lossless_opt}
\end{equation}
Substituting Eq.~\eqref{eq:lossless_opt} into Eq.~\eqref{eq:Pnet_quad} yields
\begin{equation}
\Pnet^{(2)}\big(x_{\eta,\mathrm{opt}}\big)=|t|^2+|r_{22}|^2=1
\quad\Rightarrow\quad
\eta_{\max}=1,
\label{eq:eta1}
\end{equation}
i.e., \emph{all transmitter power is delivered to the receiver side} while the auxiliary port supplies no net energy beyond shaping the interference state.

Moreover, for a lossless reciprocal two-port, unitarity also yields the constraint \cite{Pozar2012}
\begin{equation}
r_{11}t^* + t r_{22}^* = 0.
\label{eq:unitarity_constraint}
\end{equation}
Combining Eq.~\eqref{eq:unitarity_constraint} with Eq.~\eqref{eq:lossless_opt} gives $s_1^-=0$:
\begin{equation}
s_1^- = r_{11} + t\,\frac{r_{22}^*}{t^*} = r_{11} + \frac{t}{t^*}r_{22}^* =0,
\label{eq:reflection_cancel}
\end{equation}
so the transmitter experiences \emph{reflectionless delivery} through an otherwise reflective barrier.

\subsection{Beyond two ports: multi-channel barriers and wavefront control}

Real ``walls'' can couple incident energy into many channels (angles, polarizations, guided/leaky modes). A natural generalization partitions the multiport scattering matrix into left/right channel subspaces,
\begin{equation}
\begin{bmatrix}
\bm{s}_L^-\\ \bm{s}_R^-
\end{bmatrix}
=
\begin{bmatrix}
\bm{R}_L & \bm{T}_{LR}\\
\bm{T}_{RL} & \bm{R}_R
\end{bmatrix}
\begin{bmatrix}
\bm{s}_L^+\\ \bm{s}_R^+
\end{bmatrix},
\label{eq:multiport}
\end{equation}
where $\bm{s}_L^+$ is the transmitter-side incident wavefront and $\bm{s}_R^+$ is the auxiliary incident wavefront synthesized on the receiver side. If the receiver side controls enough degrees of freedom and $\bm{T}_{LR}$ is invertible over the relevant subspace, reflection cancellation on the transmitter side can be achieved by
\begin{equation}
\bm{s}_R^+ = -\bm{T}_{LR}^{-1}\,\bm{R}_L\,\bm{s}_L^+,
\label{eq:multiport_cancel}
\end{equation}
and more generally by least-squares solutions when the control is partial. This connects coherent barrier-assisted WPT to modern wavefront-shaping and transmission-matrix approaches in complex media \cite{Vellekoop2007,Popoff2010PRL,RotterGigan2017RMP,Pichler2019Nature}.

\section{Numerical and analytical demonstrations}

\subsection{Fabry--P\'erot high-index slab barrier}

\begin{figure}[t]
\centering
\includegraphics[width=\linewidth]{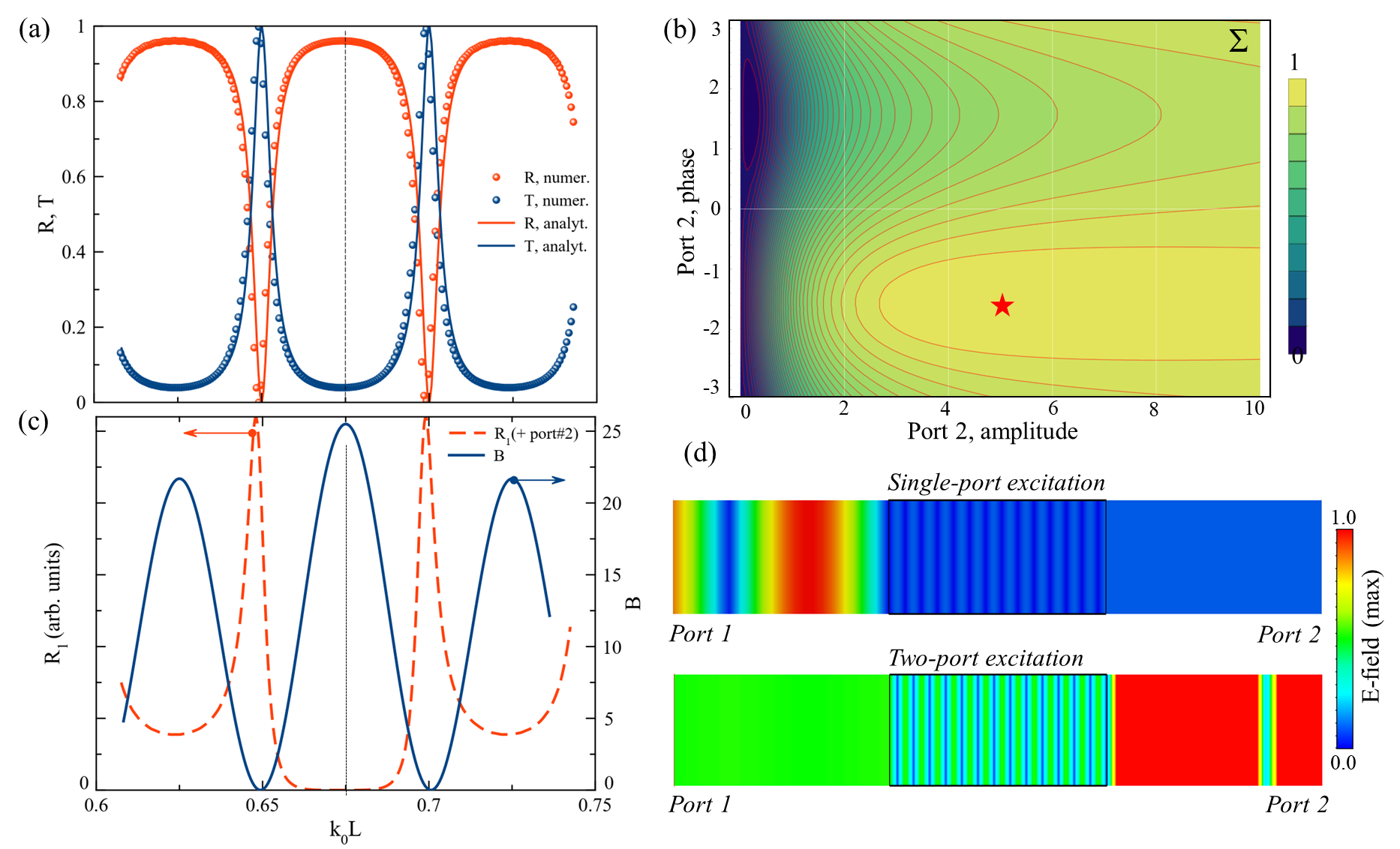}
\caption{\textbf{Fabry--P\'erot slab barrier.} (a) One-sided reflection $R$ and transmission $T$ for a dielectric slab ($\varepsilon_r=100$, $L=\SI{405}{nm}$): transfer-matrix theory (solid) and full-wave simulation (markers). (b) Fractional routing metric $\Sigma$ [Eq.~\eqref{eq:Sigma_a_phi}] versus auxiliary amplitude $a$ (horizontal) and phase $\varphi$ (vertical); star marks a near-$\Sigma=1$ operating point. (c) Enhancement $B=\Sigma/|t|^2$ and transmitter-side reflected power under coherent assistance; the optimized point yields large $B$ while suppressing reflection. (d) Field magnitude snapshots (normalized) for one-sided (top) and two-sided coherently assisted excitation (bottom), illustrating the transition from a standing-wave to a reflectionless-through state.}
\label{fig:fig3}
\end{figure}

We first consider a one-dimensional dielectric slab of thickness $L=\SI{405}{nm}$ and relative permittivity $\varepsilon_r=100$ in air (dimensionless frequency $\kz L$, $\kz=\omega/c$). The slab is a canonical reflective barrier: between Fabry--P\'erot transmission resonances, the transmission is strongly suppressed. Figure~\ref{fig:fig3}(a) shows $R=|r_{11}|^2$ and $T=|t|^2$ computed analytically via a transfer-matrix method for layered media \cite{Zhan2013} (solid curves) and verified by full-wave simulations (CST Microwave Studio; markers). In the stop region around $\kz L\approx 0.675$, the one-sided transmission drops to only a few percent while $R$ approaches unity.

Using the complex $t$ and $r_{22}$ at $\kz L=0.675$, we evaluate $\Sigma(a,\varphi)$ from Eq.~\eqref{eq:Sigma_a_phi}. Figure~\ref{fig:fig3}(b) maps $\Sigma$ versus auxiliary amplitude $a$ and phase $\varphi$. The red star denotes an operating point where $\Sigma\approx 1$, i.e., essentially all incident power from both sides exits at port~2. Importantly, the theory provides a closed-form design rule: for a lossless slab the optimal auxiliary wave is given by Eq.~\eqref{eq:lossless_opt}, which also guarantees $s_1^-=0$ (reflectionless delivery) by Eq.~\eqref{eq:reflection_cancel}.

Figure~\ref{fig:fig3}(c) reports the enhancement factor
\begin{equation}
B \equiv \frac{\Sigma}{|t|^2},
\label{eq:B_def}
\end{equation}
together with the transmitter-side reflected power in the two-sided drive configuration. Near the reflection band center, coherent assistance yields $B\sim 25$ while suppressing the reflected wave at port~1. The field distributions in Fig.~\ref{fig:fig3}(d) visualize this: one-sided excitation produces a strong standing wave, whereas the coherently assisted state exhibits a largely traveling-wave profile consistent with reflection cancellation.

\subsection{Lossy slab: receiver-side energy balance}

\begin{figure}[t]
\centering
\includegraphics[width=0.95\linewidth]{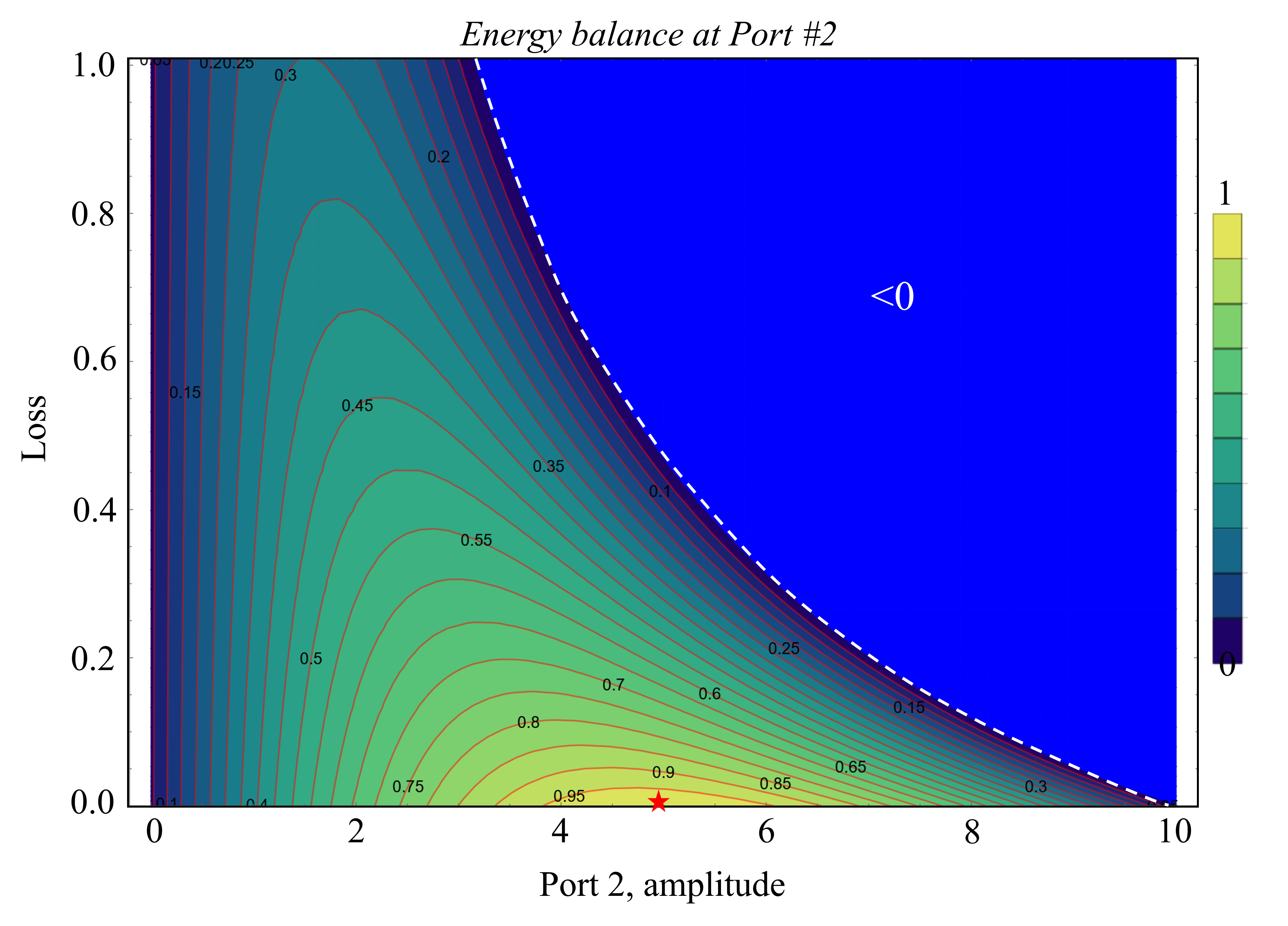}
\caption{\textbf{Lossy barrier.} Receiver-side net extracted power $\Pnet^{(2)}=|s_2^-|^2-|s_2^+|^2$ for a dielectric slab with $\varepsilon_r=100+\ii\,\mathrm{Loss}$, plotted versus loss and auxiliary amplitude $a$ (phase held near the lossless optimum). The dashed curve indicates $\Pnet^{(2)}=0$; below it the receiver extracts net power from the transmitter despite dissipation. The star marks the lossless operating point used in Fig.~\ref{fig:fig3}.}
\label{fig:fig4}
\end{figure}

Perfect $\eta=1$ is unattainable when the barrier dissipates power. To quantify usefulness in lossy systems, we report the receiver-side \emph{net} energy balance (net extracted power)
\begin{equation}
\Pnet^{(2)} = |s_2^-|^2-|s_2^+|^2,
\label{eq:energy_balance_again}
\end{equation}
which is positive when the receiver extracts more power than it injects at port~2.

Figure~\ref{fig:fig4} considers $\varepsilon_r = 100 + \ii\,\mathrm{Loss}$ and plots $\Pnet^{(2)}$ versus the loss parameter and auxiliary amplitude $a$ (with a representative phase chosen near the lossless optimum). The dashed boundary separates regimes with $\Pnet^{(2)}>0$ (useful net delivery) from those where dissipation dominates and the auxiliary port must supply net power. The key point is that substantial regions of parameter space retain positive net delivery, demonstrating that coherent assistance remains beneficial even when the barrier is not lossless. In practice, the auxiliary wave can be optimized for each frequency/loss level using Eq.~\eqref{eq:xopt_lossy} (or directly from measured $S$-parameters).

\subsection{Barely transparent perforated PEC metasurface}

\begin{figure}[t]
\centering
\includegraphics[width=\linewidth]{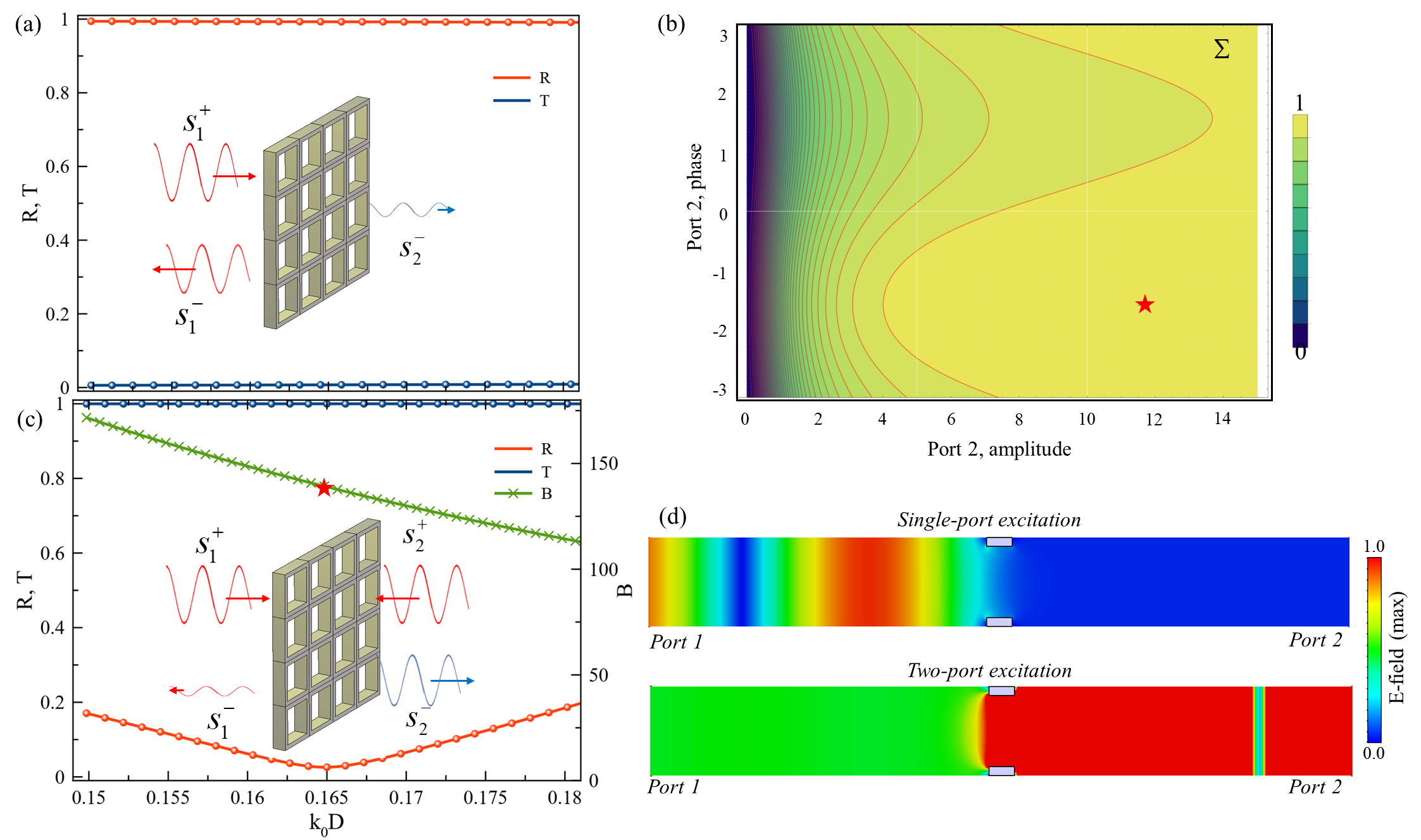}
\caption{\textbf{Perforated PEC metasurface barrier.} (a) One-sided reflection and transmission show nearly perfect reflection across the band. (b) $\Sigma(a,\varphi)$ computed from Eq.~\eqref{eq:Sigma_a_phi} at $\kz D=0.165$; star indicates an operating point with $\Sigma\approx 1$. (c) Under coherent assistance, effective transmission approaches unity and reflection is suppressed, yielding large enhancement $B$ relative to the vanishing one-sided transmission. (d) Representative field magnitude maps comparing one-sided and coherently assisted excitation, illustrating reflection suppression and forward delivery.}
\label{fig:fig5}
\end{figure}

We next study a non-resonant but strongly reflecting barrier: a PEC screen perforated by a periodic array of subwavelength rectangular holes (metasurface). Subwavelength apertures typically exhibit extremely low transmission unless tailored to support extraordinary transmission resonances \cite{Ebbesen1998,MartinMoreno2001,GarciaVidal2010}. Here the parameters are $D=\lambda_0/6$, aperture size $s=\lambda_0/7.5$, and thickness $t=\lambda_0/20$ (with representative simulated values $D=\SI{100}{nm}$, $s=\SI{80}{nm}$, $t=\SI{30}{nm}$ at $\lambda_0=\SI{600}{nm}$). Figure~\ref{fig:fig5}(a) shows that one-sided illumination yields $R\approx 1$ and $T\approx 0$ across the considered band: the barrier is essentially opaque.

Despite this, the coherent-control theory predicts that a properly phased auxiliary wave can force nearly all outgoing power to exit on the receiver side. Figure~\ref{fig:fig5}(b) maps $\Sigma(a,\varphi)$ at $\kz D=0.165$ and shows a broad region approaching $\Sigma\simeq 1$, with the star indicating one operating point. Figure~\ref{fig:fig5}(c) shows that, at the optimized operating point, the effective transmission approaches unity while reflection is strongly suppressed, yielding enhancement factors $B$ on the order of $10^2$. Field maps in Fig.~\ref{fig:fig5}(d) confirm that the coherently assisted excitation suppresses standing-wave patterns and promotes forward delivery. Importantly, the barrier itself remains unchanged; the apparent ``transparency'' is a property of the \emph{two-sided coherent drive state}.

\section{Discussion and potential applications}

The central message of this work is that an opaque barrier can become effectively transparent to a chosen transmitter--receiver pair not by altering the barrier, but by preparing an appropriate \emph{incident scattering state}. In the two-port setting, the auxiliary wave acts as a coherent ``boundary condition'' that reshapes the interference between the transmitted and reflected components. This viewpoint aligns the present mechanism with broader coherent-control paradigms in wave physics, where multi-port excitation enables qualitatively new behavior of a fixed linear system, such as coherent perfect absorption (CPA) and related ``anti-lasing'' effects \cite{Chong2010,Wan2011,Baranov2017NRM,Pichler2019Nature}. The difference is conceptual but important: CPA uses interference together with dissipation to eliminate outgoing waves, whereas here interference is used to suppress the undesired outgoing wave (back-reflection to the transmitter) and route flux toward the receiver. In this sense, the method is closer to wavefront-control and eigenchannel ideas developed for complex media, where the input state is engineered to maximize a chosen output channel \cite{RotterGigan2017RMP,Vellekoop2007,Popoff2010PRL}.

A recurring question is bandwidth. The present approach does not rely on engineering a narrow transparency resonance of the barrier (in contrast to extraordinary transmission windows in perforated screens and related structures \cite{Ebbesen1998,GarciaVidal2010}); nevertheless, the optimal auxiliary wave generally depends on frequency through the complex coefficients $t(\omega)$ and $r_{22}(\omega)$. For quasi-static barriers this frequency dependence can be handled by precompensating $a(\omega)$ and $\varphi(\omega)$ from measured S-parameters, and then generating a multi-tone or wideband auxiliary waveform whose spectrum implements the desired complex ratio (e.g., via digital predistortion and linear RF/IF processing). A complementary route is to interpret coherent assistance in the time domain: if the receiver-side node can estimate the impulse response across the barrier, time-reversal/phase-conjugation concepts show how broadband focusing and energy delivery can be achieved through complicated reverberant environments \cite{Fink1992TR,Lerosey2004TR,Lerosey2006TRWideband,Carminati2007TRCavity}. While time reversal is typically discussed for focusing and communications, it provides an instructive operational template here: measure a channel response, compute the waveform that is the time-reversed (or phase-conjugated) counterpart of the desired field at the target, and re-emit it so that energy concentrates at the receiver side despite strong multipath and scattering. This perspective is also consistent with the emerging view that programmable environments bridge classical time-reversal physics and reconfigurable intelligent surfaces (RISs) \cite{LeroseyFink2022JPROC}.

Practical realization hinges on two capabilities: (i) estimating the relevant complex scattering parameters (or a transmission matrix in the multi-channel case) and (ii) generating an auxiliary field that remains phase-coherent with the transmitter. In a controlled laboratory configuration, $t(\omega)$ and $r_{22}(\omega)$ can be acquired directly with a vector network analyzer and stable reference planes, after which the auxiliary amplitude/phase can be computed from the closed-form expressions derived in the theory section. In deployed scenarios (e.g., building walls, shielded rooms), the same information can be obtained using pilot tones and coherent channel sounding, where the receiver-side node estimates the complex field associated with transmission and reflection and then adapts $a$ and $\varphi$ in a closed loop to maximize a selected metric (e.g., received net power, reflection suppression, or a weighted combination). The most critical systems issue is maintaining phase coherence across physically separated nodes. This can be addressed with a shared frequency reference (wired reference, shared local oscillator distribution, or disciplined clocks), or through over-the-air reference dissemination and calibration. In communications, distributed coherent transmission has been demonstrated using shared-clock primitives that enable tight phase alignment among radios \cite{Abari2015AirShare}, and similar synchronization strategies can be leveraged here to maintain a stable auxiliary phase over the channel coherence time.

When the barrier couples energy into multiple propagation channels, coherent assistance becomes a wavefront-synthesis problem: the receiver side must generate not only a single complex amplitude but a controllable spatial field $\bm{s}_R^+(\omega)$ that compensates the barrier response for the chosen transmitter-side illumination. This is precisely the operating regime of wavefront shaping in complex media \cite{RotterGigan2017RMP,Vellekoop2007,Popoff2010PRL}, and it also connects naturally to RIS and programmable metasurface platforms, which provide many tunable degrees of freedom for shaping reflected (and, in some architectures, transmitted) wavefronts \cite{Pan2021RIS6G,Liang2021RIS,Bjornson2022SPM}. From an engineering standpoint, RIS-based control introduces familiar requirements: channel estimation (possibly of cascaded links), calibration, and re-optimization when the environment changes. These topics are now well developed in the RIS literature, including system-level surveys and fundamental discussions of identifiability and training overhead \cite{Swindlehurst2022JPROC,Bjornson2022SPM,Jian2022RISSurvey}. Furthermore, end-to-end programmable wireless environments have been articulated as a hardware/software stack in which metasurfaces are networked and configured by a controller that uses measurements and feedback to realize desired propagation states \cite{Liaskos2022JPROC}. In the WPT context, active or dynamically reconfigurable metasurfaces have also been experimentally explored for localizing fields and creating controllable ``hotspots'' for power delivery \cite{Ranaweera2019ActiveMSWPT}, suggesting a concrete pathway for implementing coherent assistance with spatial control rather than a single auxiliary feed.

The approach is particularly attractive when barrier modification is impractical, but receiver-side access exists to deploy an auxiliary emitter or a programmable surface. This includes shielded/ Faraday-type rooms used to suppress external interference and/or prevent emission, for which shielding effectiveness is commonly specified and verified using standard test procedures (e.g., IEEE~299) \cite{IEEE2992006}. In magnetic shielded rooms used for biomagnetism and precision sensing, the barrier is intentionally engineered to achieve extremely high shielding factors and low residual noise, which is precisely the regime where conventional one-sided RF power delivery is inhibited and penetrations are carefully controlled \cite{Bork2001MSR,Holmes2022MSR}. Coherent assistance suggests a complementary strategy: maintain the barrier’s shielding function under ordinary illumination, while enabling power delivery only when a receiver-side coherent “key” is present.

A second concrete domain is sealed vacuum and cryogenic hardware, where adding or upgrading feedthroughs is often a nontrivial cost/complexity driver and wiring itself is a major contributor to heat load and system performance limits \cite{BalshawPracticalCryogenics}. This is not merely a laboratory convenience issue: cryogenic and ultra-high-vacuum feedthroughs are the subject of dedicated engineering efforts in accelerator and instrumentation systems, where tight leak-rate, cleanliness, thermal-cycling, and RF constraints must be met \cite{Vilcins2014Feedthrough,Cianciolo2018Feedthrough}. In such platforms, coherently assisted delivery can be framed as a way to reduce reliance on additional physical penetrations for low-to-moderate power delivery to sensors, actuators, or internal electronics, or as a mechanism to “turn on” power transfer only during specific operational windows without permanently compromising the enclosure.

A third domain is powering or interrogating devices behind architectural barriers and deep-indoor locations. Building penetration loss is a recognized bottleneck in outdoor-to-indoor connectivity (and, by extension, far-field energy delivery), and is explicitly modeled in both ITU recommendations and 3GPP channel models, including frequency-dependent material loss for concrete and metallized glass \cite{TR38901,ITURP2040,ITURP2346}. In these cases, coherent assistance can be viewed as a controllable way to counteract strong reflections and poor transmission without retrofitting the wall with engineered apertures or frequency-selective structures.

Finally, the ``enabled only under coherent control'' property indicated in Fig.~\ref{fig:fig1} can be interpreted as a physical-layer access mechanism. Security, authentication, and misuse scenarios are increasingly discussed for WPT infrastructure (e.g., EV charging and other public deployments) \cite{Kumar2025SecureWPT}, and related wireless security methods such as physical-layer authentication provide a natural vocabulary for such access control \cite{Alhoraibi2023PLA}. In this context, the auxiliary coherent waveform can function as an analog authorization token: the barrier remains reflective under normal illumination, but becomes effectively transparent only when the receiver-side node emits the correct phase/amplitude pattern. Beyond pure power transfer, this naturally interfaces with simultaneous wireless information-and-power transfer (SWIPT) and joint waveform/channel design, where information and energy delivery are co-optimized \cite{Varshney2008,GroverSahai2010,ClerckxZhang2019,Cheng2024JAPmsWireless}.

The limitations of coherently assisted barrier transmission are equally clear and help frame future work. First, strict unit transmitter-to-receiver efficiency requires that the considered channel set be effectively lossless; absorption and radiation into uncontrolled channels make the scattering operator subunitary and reduce the attainable net delivery. Second, the method is sensitive to phase errors: oscillator phase noise, synchronization drift, and environmental dynamics can detune the interference condition and reduce reflection suppression. Third, in strongly multi-channel settings, achieving large gains may require many controllable degrees of freedom (multiple feeds or a spatially programmable surface) and efficient calibration/feedback protocols. These considerations do not negate the concept; rather, they identify the engineering knobs that govern performance in realistic environments and motivate hybrid strategies that combine coherent assistance with limited barrier design (e.g., modest structural measures that reduce uncontrolled leakage while keeping the barrier largely unchanged).

\section{Conclusion}

We introduced coherently assisted wireless power transfer through poorly transparent barriers using an auxiliary, phase-locked wave launched from the receiver side. A fundamental scattering formulation yields closed-form expressions for optimizing fractional routing $\Sigma$ and, crucially, the net received power $\Pnet^{(2)}$ and transmitter-to-receiver efficiency $\eta$. In the lossless two-port limit, coherent assistance cancels transmitter-side reflection and guarantees $\eta=1$ even when the barrier is nearly opaque under one-sided illumination. Analytical and full-wave examples—a high-index slab and a perforated PEC metasurface—demonstrate near-unity effective transmission and large enhancement factors without altering the barrier structure. For lossy barriers, a net-energy-balance analysis identifies regimes of positive net delivery and provides a direct route to optimize performance from measured S-parameters. Beyond WPT, the framework can be generalized to multi-channel barriers via wavefront control, suggesting applications in secure or programmable energy delivery through reflective environments.

\section*{Acknowledgments}
The author thanks colleagues and collaborators for discussions on coherent control of scattering and wireless power transfer. The author acknowledges financial support from the U.S. Department of Energy (DoE) and the U.S. Air Force Office of Scientific Research (AFOSR).

\bibliographystyle{unsrt}
\bibliography{references} 

\end{document}